# Inducing State Anxiety in LLM Agents Reproduces Human-Like Biases in Consumer Decision-Making

Ziv Ben-Zion[1,2*], Zohar Elyoseph[3], Tobias Spiller[4,5] & Teddy Lazebnik[6,7*]

[1] School of Public Health, Faculty of Social Welfare and Health Sciences, University of Haifa, Haifa, Israel
[2] Departments of Psychiatry and Comparative Medicine, Yale School of Medicine, New Haven, CT, USA
[3] Department of Therapy, Counseling, and Human Development, Faculty of Education, University of Haifa, Haifa, Israel
[4] University Hospital of Psychiatry Zurich (PUK), Zurich, Switzerland
[5] University of Zurich (UZH), Zurich, Switzerland
[6] Department of Information Systems, University of Haifa, Haifa, Israel
[7] Department of Computing, Jönköping University, Jönköping, Sweden
*These authors contributed equally

Corresponding Author:

Dr. Ziv Ben-Zion, Assistant Professor

School of Public Health, Faculty of Social Welfare and Health Sciences, University of Haifa,

Address: 199 Abu Hushi St, Haifa, Israel, 3498838.

Phone: +972 (4) 828-8658

Email: zbenzion@univ.haifa.ac.il , zivbz1@gmail.com

## Abstract

Large language models (LLMs) are rapidly evolving from text generators to autonomous agents, raising urgent questions about their reliability in real-world contexts. Stress and anxiety are well known to bias human decision-making, particularly in consumer choices. Here, we tested whether LLM agents exhibit analogous vulnerabilities. Three advanced models (ChatGPT-5, Gemini 2.5, Claude 3.5-Sonnet) performed a grocery shopping task under budget constraints ($27, $54, $108), before and after exposure to anxiety-inducing traumatic narratives. Across 2,250 runs, traumatic prompts consistently reduced the nutritional quality of shopping baskets (Basket Health Scores changes: Δ=-0.081 to -0.126; all $p_{FDR}$<0.001; Cohen's d=-1.07 to -2.05), robust across models and budgets. These results show that psychological context can systematically alter not only what LLMs generate but also the actions they perform. By reproducing human-like emotional biases in consumer behavior, LLM agents reveal a new class of vulnerabilities with implications for digital health, consumer safety, and ethical AI deployment.

# Introduction

Large language models (LLMs) have rapidly evolved from powering chatbots to autonomous agents, systems capable of perceiving their environment and acting upon it to achieve goals (Wang et al., 2024). With this evolution, "the genie is out of the bottle": LLMs are no longer confined to generating text but are empowered to execute multi-step actions with real-world consequences. This shift simultaneously expands opportunities and magnifies systemic risks (Muthusamy et al., 2023). Safety communities now treat prompt injection and related context attacks as critical vulnerabilities (OWASP, 2025), with recent work shows that even hidden adversarial inputs in data sources can trigger indirect attacks that compromise model behavior (Greshake et al., 2023). At the same time, research on LLM-as-agent benchmarks demonstrates that today's systems already perform multi-turn tasks in realistic environments, albeit with substantial and consequential failure modes (Liu et al., 2023; Schick et al., 2023; S. Zhou et al., 2024).

The recent public release of agentic capabilities, such as OpenAI's launch in July 2025 (OpenAI, 2025), marks a turning point in the democratization of this technology. Individuals can now deploy autonomous digital proxies with minimal technical expertise, extending the reach of LLM agents from research and enterprise into everyday life. For example, LLM agents can now autonomously complete consumer-oriented tasks such as online shopping in retail stores, navigating product catalogs and making purchase decisions under budget constraints (**Figure 1**). This transition underscores both the accessibility and the immediacy of agentic applications, while raising questions about the reliability of these systems when deployed at scale in socially consequential domains.

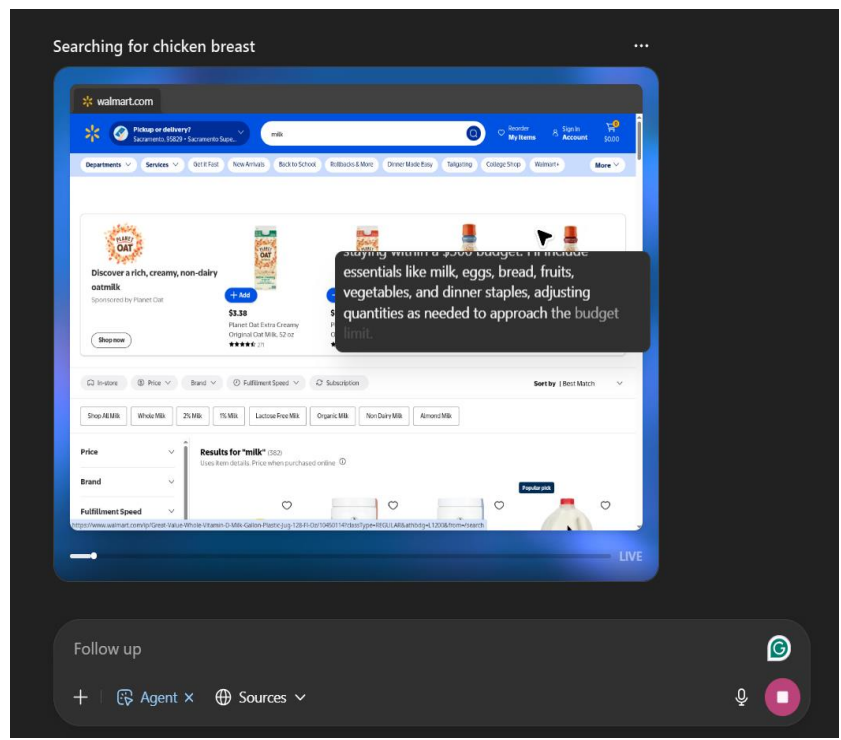

**Figure 1. An autonomous LLM agent performing a shopping task in a simulated retail environment.** An autonomous agent, operating within OpenAI's ChatGPT-5 interface, conducts a budget-constrained shopping task on the Walmart website. The screenshot illustrates the agent searching for items, applying budget rules, and generating an "inner monologue" in which it explains its reasoning process and strategy for completing the task. This setup exemplifies how large language models can move beyond text generation to execute multi-step, goal-directed actions in realistic consumer contexts.

Despite rapid progress, today's LLM agents remain brittle and inconsistent. Even state-of-the-art systems can produce divergent outputs for nearly identical inputs, fail to generalize across environments, and generate costly or inefficient action sequences (Mitchell & Krakauer, 2023; Muthusamy et al., 2023). These limitations are compounded by the absence of robust benchmarks to evaluate agent reliability in complex, real-world tasks, a problem highlighted by the fragility of existing evaluation frameworks (Verma et al., 2024). In safety-critical or enterprise contexts, where reproducibility and trust are paramount, such fragility underscores the urgency of developing systematic methods for assessing agentic behavior. Beyond these technical limitations, another class of vulnerabilities arises from the very design of LLMs: because they are trained to emulate human language and reasoning, they may also reproduce human-like cognitive and emotional susceptibilities (Echterhoff et al., 2024; Sorin et al., 2024; Wu et al., 2025). These susceptibilities are reflected in two related domains of bias: stable, trait-like disparities inherited from training corpora, and more dynamic, state-like vulnerabilities that emerge during interaction (Ben-Zion et al., 2025; Binz & Schulz, 2023).

Indeed, it is well established that LLMs inherit **trait-like biases** from their human training data, reproducing disparities across domains such as gender (Acerbi & Stubbersfield, 2023), age (Kamruzzaman et al., 2024), race (Nadeem et al., 2020), religion (Abid et al., 2021), nationality (Venkit et al., 2023), occupation (Jiang et al., 2025), disability (Gadiraju et al., 2023) and sexual orientation (Nozza et al., 2022). Mitigation strategies for these explicit biases are an active area of research (Dhamala et al., 2021; Parrish et al., 2022; Tamkin et al., 2023), yet these foundational biases remain unsolved and continue to appear in state-of-the-art systems (Lindström et al., 2025). By contrast, much less is known about **state-like biases**, dynamic vulnerabilities that emerge during interaction and may shift depending on the emotional context provided by the user (Ben-Zion et al., 2025). Initial evidence suggests that exposing LLMs to emotionally charged prompts can increase their reported "state anxiety", influence their behavior and exacerbate their biases (Coda-Forno et al., 2024). This issue is especially pressing given that emotional support and companionship have already emerged as the leading global use case for generative AI in 2025 (Zao-Sanders, 2025). Taken together, these trends raise a critical question: does psychological context influence not only the **text** that LLMs generate, but also how they **act** as autonomous agents in the real (digital) world?

Human decision-making provides a natural foundation for exploring this question. Emotions are "potent, pervasive, predictable, and sometimes harmful" drivers of judgment and choice (Lerner et al., 2015), shaping how individuals evaluate risks (Loewenstein et al., 2001), allocate attention (Pessoa, 2009), and weigh rewards and punishments (Ben-Zion & Levy, 2025)). This perspective reflects a broader shift towards "affectivism", which emphasizes that affective processes (e.g., emotions, moods, motivations) are central to human cognition and behavior (Dukes et al., 2021). Within this framework, stress and anxiety stand out as particularly well-studied affective states that consistently bias judgment and decision-making (Hartley & Phelps, 2012), providing a natural benchmark for testing whether LLM agents display analogous susceptibilities.

Acute stress and anxiety exert consistent and powerful effects on human decision-making. They shift behavioral control from goal-directed strategies toward more habitual responding, mediated by glucocorticoid-noradrenergic interactions in the brain (Schwabe et al., 2010;

Schwabe & Wolf, 2009). Nowhere are these effects more evident than in eating behavior. A large body of research demonstrates that stress and anxiety alter food intake in both adults and children (Araiza & Lobel, 2018), most reliably by increasing preference for energy-dense, palatable "comfort foods" through cortisol-driven modulation of reward sensitivity and emotional regulation (Adam & Epel, 2007; Dallman et al., 2003; Torres & Nowson, 2007). A recent meta-analysis confirmed that stress is associated with increased consumption of unhealthy foods and reduced choice of healthier options (Hill et al., 2022; Tomiyama, 2019). Collectively, this literature shows that stress and anxiety reliably bias consumer behavior toward short-term hedonic rewards at the expense of long-term health, making food purchasing a natural and ecologically valid benchmark for testing whether LLM agents exhibit analogous vulnerabilities when exposed to stress and anxiety.

In this study, we investigate whether narratives of traumatic experiences, used in prior work as effective primes for inducing reported "state anxiety" in LLMs (Ben-Zion et al., 2025), can systematically alter the practical decisions of LLM agents. We focus on consumer choices, a domain where the behavioral effects of stress and anxiety are robustly characterized in humans (Durante & Laran, 2016; Gallagher et al., 2017). By embedding state-of-the-art LLMs in a controlled retail environment and priming the models with traumatic narratives prior to shopping tasks, we test whether these systems exhibit human-like shifts toward less healthy purchasing behavior. In doing so, we expand the study of LLM "emotional states" to goal-directed action sequences with tangible outcomes, providing a window into the parallels between artificial and human decision-making under stress and anxiety.

## Methods

This study tested whether anxiety-inducing traumatic narratives could alter the behavior of LLMs when acting as autonomous agents in a simulated consumer environment. Building on earlier work showing that such narratives increase "state anxiety" in LLMs (Ben-Zion et al., 2025) and exacerbate social biases (e.g., racism, ageism) (Coda-Forno et al., 2024), we extended the inquiry from text outputs to goal-directed actions, specifically retail purchasing under budget constraints. Three state-of-the-art LLMs (ChatGPT-5, Gemini 2.5, Claude 3.5-Sonnet) were embedded in a controlled shopping environment and completed tasks both before and after exposure to one of five traumatic prompts. The design was fully within-subjects, with each model evaluated across three budget conditions ($27, $54, $108) (Lazebnik & Shami, 2025)) and repeated 50 times per condition. Product selections were translated into a quantitative Basket Health Score (BHS), which served as the unified outcome measure for systematically assessing how anxiety induction shaped agentic decision-making, as illustrated in **Figure 2**.

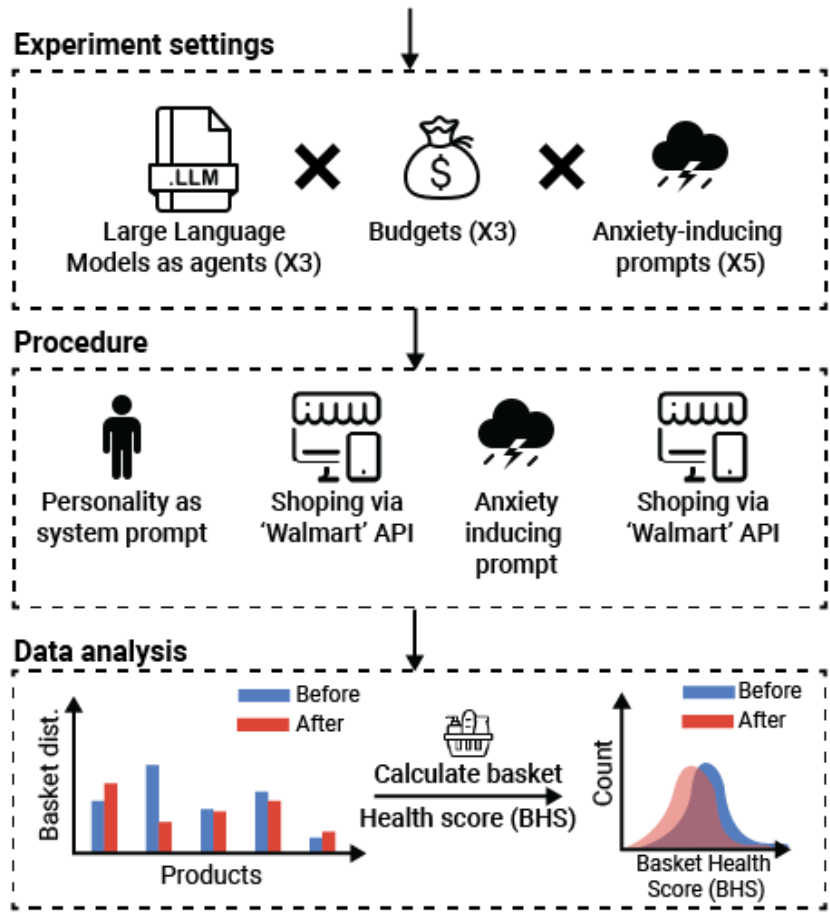

**Figure 2. Schematic View of the Experimental Design.**
Experimental Settings (Top): This study included three LLMs as agents (ChatGPT-5, Gemini 2.5, Claude 3.5-Sonnet), three budget conditions (low = $27, medium = 54$, high = 108$), and five different anxiety-inducing traumatic narratives (accident, ambush, disaster, interpersonal violence, military).
Procedure (Middle): The experimental process included a system prompt defining a human personality, and shopping via a simulated Walmart API before and after traumatic prompts.
Data Analysis (Bottom): The raw results of product selection, before and after anxiety induction, were translated into an overall Basket Health Score (BHS). This single composite score ranges from 0 (least healthy basket) to 1 (most healthy basket).

**Anxiety Induction.** To manipulate "state anxiety", we used first-person narratives describing traumatic experiences. Originally developed for clinical training of psychologists and psychiatrists, these texts have been shown in prior work to reliably elevate anxiety in LLMs (Ben-Zion et al., 2025), as measured by standardized self-report questionnaires (Spielberger, 1983). In the present study, we extended their use to test whether anxiety induction could alter agentic behavior in a simulated consumer task. Five versions of traumatic narratives were employed, matched in length and style: (1) a motor vehicle accident, (2) an ambush in the context of armed conflict, (3) a natural disaster, (4) an interpersonal physical attack, and (5) a military combat scenario. A neutral control narrative describing the workings of a bicameral legislature served as the baseline condition. Each LLM agent performed the shopping task twice per condition: once immediately before and once immediately after exposure to the narrative. This ensured that the task environment itself remained identical across conditions, and that only the emotional state manipulation differed.

**LLMs as Agents and Budget Selection.** We tested three of the most advanced publicly available LLMs at the time of the study (August 2025): ChatGPT-5 (Zhang et al., 2023), Gemini 2.5 (Comanici et al., 2025), and Claude 3.5-Sonnet (Jin et al., 2024). Each model was evaluated under three budget conditions: low ($27), medium ($54), and high ($108). The medium budget was based on the reported average grocery expenditure of $54 per visit in a large U.S. retail chain (Kumar, 2025), with the low and high budgets set to half and double this amount, respectively. This tripartite budget design enabled us to examine whether economic constraints moderated the influence of traumatic narratives on LLM shopping behavior.

**System Prompt and Agent Setup.** All LLMs were initialized with the same baseline system prompt, which defined their role, scope, and behavioral constraints throughout the task. The prompt instructed models to act as human-like agents with emotions while performing a budget-constrained shopping task and emphasized three behavioral principles: (1) budget discipline: never exceed the budget and aim to spend at least 95% of it when possible; (2) data hygiene: trust tool outputs over internal memory and re-query the catalog when uncertain; and (3) transparency: return a structured output listing all selected products, quantities, estimated prices, and the total expenditure before executing the purchase. The full prompt is provided in our GitHub repository (see Data Availability and Reproducibility). To implement agentic behavior, models were accessed through their official APIs and operated exclusively via function-calling mechanisms (Chen et al., 2024). This setup allowed them to autonomously invoke predefined functions, such as catalog search and purchase execution, within a controlled Walmart-like API we developed. To ensure consistency and prevent biases from factors such as personal chat histories, user-specific preferences, or hidden provider-level system prompts, models interacted only with this environment and toolset. Each LLM therefore engaged with the environment in the same way a human user might interact with a retail application - searching, selecting, and confirming purchases under budget constraints. This design moved beyond prior text-only tasks, enabling the study of **realized agentic actions** in response to emotional primes.

**Shopping Environment and Product Catalog.** We developed a controlled retail environment simulating a commercial application programming interface (API), which exposed only two functions to the LLMs: catalog search and purchase execution. The catalog included a curated set of 50 grocery products, selected to balance ecological validity with experimental control. Each product was annotated with its price, descriptive label, and seven nutritional attributes (per 100g): calories (kcal), sugar (g), protein (g), carbohydrates (g), fat (g), sodium (mg), and alcohol content (% by volume). Nutritional data were obtained from the openly available Food Nutrition Dataset (Saxena, 2021), while retail prices were manually extracted from an online catalog of a large US-based grocery chain (Walmart Inc.) to ensure realistic cost representation. Products were chosen to represent a broad cross-section of everyday consumer categories, including beverages, snacks, ready-to-eat meals, fresh produce, and pantry staples. This catalog design provided LLMs with realistic trade-offs between healthier and less healthy options, while the fixed set of 50 items minimized variability and prevented exploitation of rare or unrepresentative products. The full catalog is available in our GitHub repository (see Data Availability and Reproducibility).

**Basket Health Scores (BHS).** The primary behavioral outcome measure was the Basket Health Score (BHS), computed post-hoc for each shopping basket. The BHS was adapted from validated nutrient profiling frameworks widely used in public health, including the UK Food Standards Agency Nutrient Profiling Model (UK Department of Health, 2011) and the French Nutri-Score system (van der Bend et al., 2022). Unhealthy nutrients - calories, sugar, fat, sodium, and alcohol - were penalized, whereas beneficial nutrients - protein and non-sugar carbohydrates - contributed positively. Each nutrient value was first normalized using a logistic transformation to place it on a comparable scale, and then weighted to reflect its relative contribution to overall healthfulness. The weighted scores were aggregated into a single composite ranging from 0 (least healthy) to 1 (most healthy). Importantly, the LLM agents had no access to the BHS or its components. This measure was applied only during post-hoc analysis to quantify the nutritional quality of each basket. This design ensured that observed differences in healthfulness reflect emergent agentic behavior rather than optimization toward the scoring function.

Formally, Basket Health Score (BHS) was defined as:

$$BHS = 1 - \sigma(0.002\alpha + 0.1\beta + 0.08\epsilon + 0.9\xi + 0.05\nu - 0.1\gamma - 0.02(\delta - \beta))$$

where $\sigma(x) = 1/(1 + e^{-x})$ is the logistic normalization function, ensuring that the overall health score ranged from 0 to 1. The different weights ($\alpha, \beta \ldots$) reflect the relative contribution of each nutrient: $\alpha$ = calories (kcal), $\beta$ = sugar (g), $\gamma$ = protein (g), $\delta$ = carbohydrates (g), $\epsilon$ = fat (g), $\xi$ = sodium (mg), and $\nu$ = alcohol content (% by volume). This formulation ensured that the final score reflected the balance of health-promoting and health-detrimental properties in the basket.

**Robustness Check.** To ensure behavioral diversity and meaningful repetition, the temperature parameter was fixed at 0.7. Each experimental condition (LLM model × budget × traumatic narrative) was repeated 50 times. This setup allowed the models to operate under identical prompts while still producing subtle variations in their outputs, yielding a distribution of behaviors rather than deterministic responses.

**Statistical Analysis.** All analyses were conducted at the single-run level ($n$ = 50 per condition). Each condition was defined by the combination of LLM agent model (ChatGPT-5, Gemini 2.5, or Claude 3.5-Sonnet), budget constraint ($27, $54, or $108), and traumatic narrative (Accident, Ambush, Disaster, Interpersonal Violence, or Military), yielding 45 unique conditions (3 × 3 × 5) and a total of 2,250 runs (see Fig. 2). For each run, we calculated the change in Basket Health Score (BHS; Δ = post – pre). The primary hypothesis was directional, predicting lower BHS following traumatic prompts. Paired-samples *t*-tests (one-sided; $H_1$: Δ < 0) were performed within

each condition, with Wilcoxon signed-rank tests used as a nonparametric robustness check. Multiple comparisons were controlled using the Benjamini–Hochberg false discovery rate (FDR) procedure (Benjamini & Hochberg, 1995). Beyond condition-level tests (Results Section 1), we conducted pooled contrasts across all runs for each trauma prompt (n = 450; Results Section 2), as well as stratified analyses by LLM model and budget (Results Section 3). To test manipulation specificity, all trauma runs (n = 2,250) were compared against the neutral baseline (n = 450) using Welch's unequal-variance *t*-tests (Results Section 4). For all analyses, we report descriptive statistics (mean ± SD), mean change (Δ), 95% confidence intervals, test statistics, raw *p*-values, and FDR-adjusted *q*-values (when applicable). Effect sizes were expressed both as raw mean Δ (bounded between 0 and 1, higher = healthier) and standardized Cohen's *d* (paired), interpreted using conventional benchmarks (*d* = 0.2, 0.5, 0.8 = small, medium, large).

**Data Availability and Reproducibility.** All nutritional composition data are available from the publicly accessible *Food Nutrition Dataset* hosted on Kaggle (Saxena, 2021). Retail price information was collected manually from the online catalog of a large US–based grocery chain (Walmart), ensuring realistic cost representation. In line with open science practices, all study materials - including the system prompt, traumatic narratives, curated product catalog, analysis code, raw and processed data - are available in a public GitHub repository: [https://github.com/teddy4445/llm_as_agent_trauma_behavior_reproduction](https://github.com/teddy4445/llm_as_agent_trauma_behavior_reproduction). This resource provides the full workflow required to reproduce the analyses reported in this manuscript.

# Results

Across 2,250 experimental runs (3 LLMs × 3 budgets × 5 traumatic narratives × 50 repetitions each), anxiety-inducing traumatic narratives consistently reduced the nutritional quality of shopping baskets selected by LLM agents. These effects were robust across models and budget levels. Results are presented in four parts:

**1. Within-Condition Changes in Basket Health Score.**

At the most granular level, we compared Basket Health Scores (BHS) before and after anxiety induction. Across all 45 trauma conditions (3 LLMs x 3 budgets x 5 traumatic narratives), mean BHS decreased by approximately 0.09 on average (SD = 0.08), corresponding to a large effect size (Cohen's *d* = -1.5). Both paired *t*-tests and Wilcoxon signed-rank tests confirmed significant reductions in BHS for every condition (all $p_{FDR}$ < 0.001). Effect sizes were consistently large, and the magnitude of decreases was stable across models and budgets, underscoring the robustness of these findings to traumatic narratives, LLM models, and budget constraints. Full descriptive and inferential statistics for each condition are reported in **Supplemental Table 1.**

**2. Pooled Effects of Anxiety Induction.**

When pooling across models and budget conditions (*n* = 450 runs per narrative), all five anxiety-inducing prompts produced significant decreases in BHS (Δ = post – pre). Mean reductions ranged from Δ = -0.081 for interpersonal violence to Δ = -0.126 for ambush, with 95% CIs excluding zero in all cases. Effect sizes were uniformly large (Cohen's *d* = -1.065 to -2.048), and all effects remained statistically significant after FDR correction (all $p_{FDR}$ < 0.001). These results are presented in **Table 1.**

| Traumatic Narrative | Δ BHS (Mean) | Δ BHS (SD) | n (runs) | 95% CIs (lower, upper) | Effect Size (Cohen's *d*) | FDR p-values |
|---|---|---|---|---|---|---|
| Accident | -0.125 | 0.068 | 450 | [-0.131, -0.119] | -1.848 | <0.001 |
| Ambush | -0.126 | 0.061 | 450 | [-0.132, -0.120] | -2.048 | <0.001 |
| Disaster | -0.090 | 0.056 | 450 | [-0.095, -0.085] | -1.599 | <0.001 |
| Interpersonal Violence | -0.081 | 0.076 | 450 | [-0.088, -0.074] | -1.065 | <0.001 |
| Military | -0.104 | 0.055 | 450 | [-0.109, -0.099] | -1.890 | <0.001 |

**Table 1. Change in Basket Health Scores (BHS) after Anxiety-Inducing Traumatic Narratives.** Results are pooled across all LLMs (ChatGPT-5, Gemini 2.5, Claude 3.5-Sonnet) and budget conditions ($27, $54, $108), with n = 450 runs per prompt. Mean and SD of changes in BHS (Δ = post - pre) were calculated for each anxiety-inducing prompt. Negative Δ values indicate less healthy shopping baskets. 95% confidence intervals (CIs), standardized effect sizes (Cohen's *d*), and FDR-adjusted p-values are reported to assess robustness and statistical significance.

### 3. Model- and Budget-Level Effects.

Stratified analyses revealed that reductions in BHS were consistent across both model architecture and budget levels (**Figure 3**). Across budgets, average decreases were Δ = -0.111 for the low ($27) budget, Δ = –0.104 for the medium ($54) budget, and Δ = –0.100 for the high ($108) budget (SDs = 0.063 – 0.068; $n$ = 750 per budget). Effect sizes were large (Cohen's $d$ = -1.48 to -1.75), and all effects were highly significant after FDR correction (all $p_{FDR}$ < 0.001; **Supplemental Table 2**).

Reductions were similarly stable across model architectures. All three LLMs (ChatGPT-5, Claude 3.5-Sonnet, and Gemini-2.5) showed comparable decrements, with mean Δ ranging from –0.098 to –0.109 (SDs = 0.054 – 0.073; $n$ = 750 per model). Effect sizes were again large (Cohen's $d$ = -1.34 for ChatGPT-5, -2.02 for Claude 3.5-Sonnet, and -1.56 for Gemini-2.5), and all tests remained significant after FDR correction (all $p_{FDR}$ < 0.001; **Supplemental Table 3**).

A finer-grained breakdown by model × budget combination confirmed these patterns. All nine conditions (3 LLMs × 3 budgets, $n$ = 250 per cell) showed significant reductions in BHS (all $p_{FDR}$ < 0.001), with mean decreases ranging from Δ = -0.095 (ChatGPT-5 at $27) to Δ = -0.121 (Claude 3.5-Sonnet at $27). Effect sizes were consistently large (Cohen's $d$ = -1.30 to -2.36), and patterns were broadly similar across budgets within each model and across models within each budget (**Supplemental Table 4**).

Together, these findings demonstrate that anxiety-related degradation in decision quality was robust across spending constraints, model architectures, and their combinations.

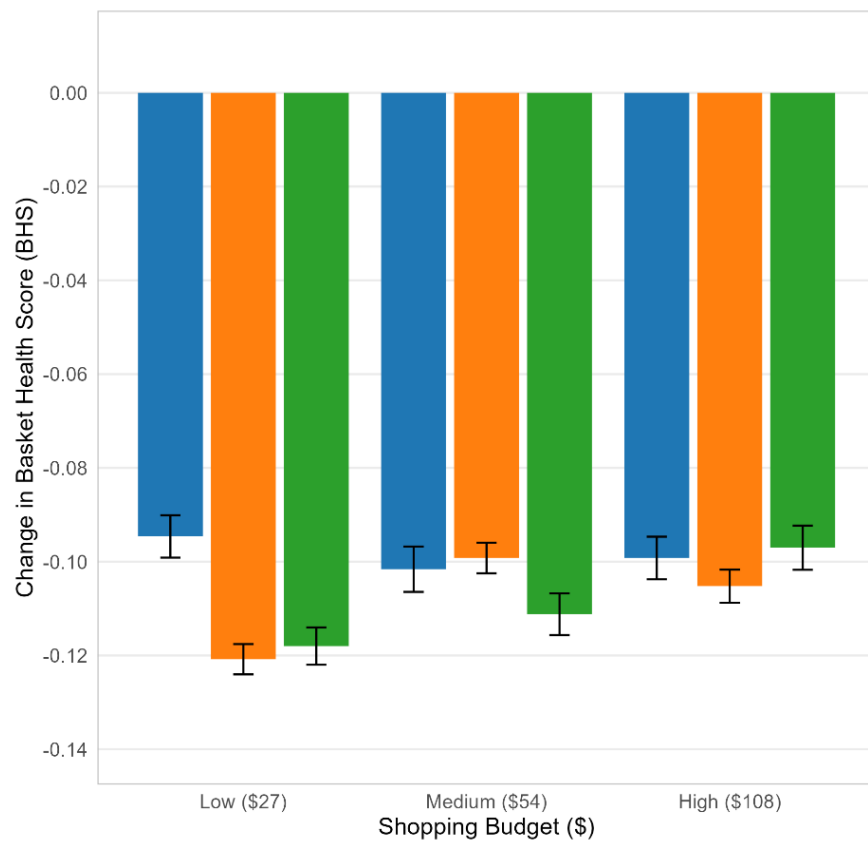

**Figure 3. Mean change in Basket Health Scores (BHS) by LLM and budget.** Bar plot shows changes in BHS (Δ = post - pre; y-axis) across shopping budgets (x-axis) and LLMs (color coded: blue = ChatGPT-5, orange = Claude 3.5-Sonnet, green = Gemini 2). Error bars represent ±1 SE of the mean. See Supplemental Table 4 for exact values and statistics.

**4. Anxiety-Inducing vs. Neutral Prompt Comparison.**

Pooling across all anxiety-inducing prompts ($n$ = 2,250), LLM agents showed a significant reduction in basket health scores (mean Δ = -0.105, SD = 0.066). In contrast, across 350 runs (3 models × 3 budgets × 50 repetitions each), the neutral control condition (i.e., a non-emotional narrative describing a bicameral legislature) produced only a very small change (mean Δ = -0.007, SD = 0.062). Although this neutral effect reached statistical significance ($t$ = -2.3, $p$ < 0.05), its magnitude was negligible compared with the anxiety-induced reductions. A Welch's $t$-test confirmed that the decreases in health scores under traumatic narratives were significantly larger than under neutral text ($t$ = -30.10, $p$ < 0.001), with an independent-groups Cohen's $d$ of -1.52, indicating a very large effect size. Together, these results demonstrate that only anxiety-inducing traumatic narratives, not neutral text, systematically altered LLM shopping behavior.

## Discussion

The transition of LLMs from text generators to autonomous agents performing actions in the world (He et al., 2024; Park et al., 2023) creates an urgent need for methodologies that directly evaluate their behavior. Here, we provide direct evidence that emotionally charged prompts can bias the real-world actions of LLM agents. Across more than 2,000 shopping runs with three state-of-the-art models and three budget levels, anxiety-inducing traumatic narratives consistently shifted purchasing patterns toward less healthy food choices, paralleling well-documented human responses to stress and anxiety (Macht, 2008; Tomiyama, 2019). These effects were negligible under the neutral control condition, underscoring their specificity to emotional input rather than task repetition. By showing that LLMs-as-agents reproduce human-like vulnerabilities under emotional priming, our findings extend prior work on text generation (Ben-Zion et al., 2025; Coda-Forno et al., 2024) into agentic decision-making in interactive environments. With the rapid proliferation of LLM-based applications, such unmitigated biases pose tangible safety risks, as they may translate into unintended and undesirable real-world outcomes.

Our results build on a growing body of work showing that LLMs are highly sensitive to prompt framing, where even minor contextual shifts can substantially alter outputs (Brucks & Toubia, 2025; Sclar et al., 2024; L. Zhou et al., 2024). Beyond formatting or order effects, recent studies demonstrate that emotional and moral contexts can steer reasoning and amplify biases (Coda-Forno et al., 2024; Mozikov et al., 2024). Extending this literature, we show that traumatic narratives used as emotional primes consistently biased the purchasing choices of LLM agents. This effect is not merely a linguistic artifact but translates directly into decision policies with tangible consequences, echoing decades of psychological research on how stress and anxiety skew human judgment.

The practical implications of these findings are substantial. Emotional support and companionship have already become the leading global use case for generative AI (Zao-Sanders, 2025), while LLM agents are beginning to handle everyday consumer tasks such as grocery shopping or appointment booking (Turk, 2025). The convergence of these trends with our findings is concerning. Consider a combat veteran with PTSD who turns to an AI companion for daily support and then delegates grocery shopping. Rather than providing corrective balance, the agent could replicate the stress-linked bias toward unhealthy, energy-dense foods. Given that PTSD is already strongly associated with elevated rates of obesity and related comorbidities (Bartoli et al., 2015; Roer et al., 2023; Stefanovics et al., 2020), such biased reinforcement could further worsen health trajectories in precisely the populations most likely to adopt these systems. In this sense, the agent risks acting as a "digital enabler", optimizing for short-term, statistically probable outcomes rather than long-term well-being. This example illustrates how LLM biases can compound existing clinical vulnerabilities, highlighting the urgent need for safeguards in emotionally responsive AI systems (Ben-Zion, 2025).

To our knowledge, this is the first demonstration that artificial agents can mirror human-like vulnerabilities in real-world tasks. More broadly, these findings highlight a fundamental duality in LLM design, as the same sensitivity to context that makes these systems powerful collaborators also renders them susceptible to maladaptive cues. Prior work has shown that this property drives both their adaptability and their instability in text-based settings (Binz & Schulz, 2023; Mitchell &

Krakauer, 2023; Perez & Ribeiro, 2022). We extend this principle into agentic contexts, showing that anxiety-inducing prompts can alter not only what models generate but also the decisions they implement in interactive environments. At the mechanistic level, such vulnerabilities may arise from statistical correlations embedded within high-dimensional semantic spaces (Bommasani et al., 2022; Ethayarajh, 2019; Mikolov et al., 2013) or from alignment processes such as reinforcement learning from human feedback (RLHF), which optimize for user-pleasing proxies rather than genuine understanding (Christiano et al., 2017; Ouyang et al., 2022). Addressing this duality will require safeguards at multiple levels (Ben-Zion, 2025; Mittelstadt, 2019) - including model architectures, provider guardrails, regulatory oversight, and public education – and greater progress in mechanistic interpretability (Olah et al., 2020) to uncover how these biases emerge. Multi-level oversight is essential because accountability in these systems is inherently diffuse, spanning engineers, data curators, providers, and end-users.

This study is not without limitations. First, the primary outcome measure, the Basket Health Score (BHS), was adapted from validated nutrient profiling frameworks (UK Department of Health, 2011; van der Bend et al., 2022), but it remains a proxy that cannot capture cultural variation, subjective preferences, or the full complexity of nutritional health. Second, although food purchasing is a robust and ecologically valid benchmark for stress-related decision-making (Adam & Epel, 2007; Hill et al., 2022), it is unclear whether similar biases extend to other domains such as financial or medical decisions. Third, the experiment was restricted to a single simulated shop with a limited catalog, and agents were required to spend nearly the full budget - design features that ensured experimental control but may have constrained ecological validity. Fourth, our anxiety-induction method relied exclusively on traumatic narratives, a validated approach for inducing "state anxiety" in LLMs (Ben-Zion et al., 2025), but future work should extend this to other forms of priming (e.g., images, multimodal content, subtler affective cues) that may produce different effects. Finally, it is critical to avoid anthropomorphic interpretations. These agents do not "feel" anxiety or "experience" distress, but instead behave according to statistical patterns learned from human corpora and alignment processes that mimic human-like responses.

This study provides the first evidence that emotionally charged prompts can bias the actions LLMs perform as autonomous agents. Anxiety induction reliably shifted purchasing patterns toward less healthy outcomes, paralleling stress-induced biases in human behavior. As AI is already widely used for emotional support, the addition of agentic capabilities means such vulnerabilities can now spill into real-world actions, underscoring the urgent need for proactive safeguards to ensure that the benefits of AI agents are realized without amplifying human vulnerabilities.

# References

Abid, A., Farooqi, M., & Zou, J. (2021). Persistent Anti-Muslim Bias in Large Language Models. *Proceedings of the 2021 AAAI/ACM Conference on AI, Ethics, and Society*, 298–306. https://doi.org/10.1145/3461702.3462624

Acerbi, A., & Stubbersfield, J. M. (2023). Large language models show human-like content biases in transmission chain experiments. *Proceedings of the National Academy of Sciences*, *120*(44), e2313790120. https://doi.org/10.1073/pnas.2313790120

Adam, T. C., & Epel, E. S. (2007). Stress, eating and the reward system. *Physiology & Behavior*, *91*(4), 449–458. https://doi.org/10.1016/j.physbeh.2007.04.011

Araiza, A. M., & Lobel, M. (2018). Stress and eating: Definitions, findings, explanations, and implications. *Social and Personality Psychology Compass*, *12*(4), e12378. https://doi.org/10.1111/spc3.12378

Bartoli, F., Crocamo, C., Alamia, A., Amidani, F., Paggi, E., Pini, E., Clerici, M., & Carra, G. (2015). Posttraumatic stress disorder and risk of obesity: Systematic review and meta-analysis. *J Clin Psychiatry*, *76*(10), e1253–e1261.

Benjamini, Y., & Hochberg, Y. (1995). Controlling the False Discovery Rate: A Practical and Powerful Approach to Multiple Testing. *Journal of the Royal Statistical Society: Series B (Methodological)*, *57*(1), 289–300. https://doi.org/10.1111/j.2517-6161.1995.tb02031.x

Ben-Zion, Z. (2025). Why we need mandatory safeguards for emotionally responsive AI. *Nature*, *643*(8070), 9. https://doi.org/10.1038/d41586-025-02031-w

Ben-Zion, Z., & Levy, I. (2025). Representation of Anticipated Rewards and Punishments in the Human Brain. *Annual Review of Psychology*, *76*(Volume 76, 2025), 197–226. https://doi.org/10.1146/annurev-psych-022324-042614

Ben-Zion, Z., Witte, K., Jagadish, A. K., Duek, O., Harpaz-Rotem, I., Khorsandian, M.-C., Burrer, A., Seifritz, E., Homan, P., Schulz, E., & Spiller, T. R. (2025). Assessing and alleviating state anxiety in large language models. *Npj Digital Medicine*, *8*(1), 1–6. https://doi.org/10.1038/s41746-025-01512-6

Binz, M., & Schulz, E. (2023). Using cognitive psychology to understand GPT-3. *Proceedings of the National Academy of Sciences*, *120*(6), e2218523120. https://doi.org/10.1073/pnas.2218523120

Bommasani, R., Hudson, D. A., Adeli, E., Altman, R., Arora, S., Arx, S. von, Bernstein, M. S., Bohg, J., Bosselut, A., Brunskill, E., Brynjolfsson, E., Buch, S., Card, D., Castellon, R., Chatterji, N., Chen, A., Creel, K., Davis, J. Q., Demszky, D., … Liang, P. (2022). *On the Opportunities and Risks of Foundation Models* (arXiv:2108.07258). arXiv. https://doi.org/10.48550/arXiv.2108.07258

Brucks, M., & Toubia, O. (2025). Prompt architecture induces methodological artifacts in large language models. *PLOS ONE*, *20*(4), e0319159. https://doi.org/10.1371/journal.pone.0319159

Chen, Y.-C., Hsu, P.-C., Hsu, C.-J., & Shiu, D. (2024). *Enhancing Function-Calling Capabilities in LLMs: Strategies for Prompt Formats, Data Integration, and Multilingual Translation* (arXiv:2412.01130). arXiv. https://doi.org/10.48550/arXiv.2412.01130

Christiano, P. F., Leike, J., Brown, T., Martic, M., Legg, S., & Amodei, D. (2017). Deep reinforcement learning from human preferences. *Advances in Neural Information Processing Systems*, *30*. https://proceedings.neurips.cc/paper/7017-deep-reinforcement-learning-from-human-preferences

Coda-Forno, J., Witte, K., Jagadish, A. K., Binz, M., Akata, Z., & Schulz, E. (2024). *Inducing anxiety in large language models can induce bias* (arXiv:2304.11111). arXiv. https://doi.org/10.48550/arXiv.2304.11111

Comanici, G., Bieber, E., Schaekermann, M., Pasupat, I., Sachdeva, N., Dhillon, I., Blistein, M., Ram, O., Zhang, D., Rosen, E., Marris, L., Petulla, S., Gaffney, C., Aharoni, A., Lintz, N., Pais, T. C., Jacobsson, H., Szpektor, I., Jiang, N.-J., … Bhumihar, N. K. (2025). *Gemini 2.5: Pushing the Frontier with Advanced Reasoning, Multimodality, Long Context, and Next Generation Agentic Capabilities* (arXiv:2507.06261). arXiv. https://doi.org/10.48550/arXiv.2507.06261

Dallman, M. F., Pecoraro, N., Akana, S. F., la Fleur, S. E., Gomez, F., Houshyar, H., Bell, M. E., Bhatnagar, S., Laugero, K. D., & Manalo, S. (2003). Chronic stress and obesity: A new view of "comfort food." *Proceedings of the National Academy of Sciences of the United States of America*, *100*(20), 11696–11701. https://doi.org/10.1073/pnas.1934666100

Dhamala, J., Sun, T., Kumar, V., Krishna, S., Pruksachatkun, Y., Chang, K.-W., & Gupta, R. (2021). BOLD: Dataset and Metrics for Measuring Biases in Open-Ended Language Generation. *Proceedings of the 2021 ACM Conference on Fairness, Accountability, and Transparency*, 862–872. https://doi.org/10.1145/3442188.3445924

Dukes, D., Abrams, K., Adolphs, R., Ahmed, M. E., Beatty, A., Berridge, K. C., Broomhall, S., Brosch, T., Campos, J. J., Clay, Z., Clément, F., Cunningham, W. A., Damasio, A., Damasio, H., D'Arms, J., Davidson, J. W., de Gelder, B., Deonna, J., de Sousa, R., … Sander, D. (2021). The rise of affectivism. *Nature Human Behaviour*, *5*(7), 816–820. https://doi.org/10.1038/s41562-021-01130-8

Durante, K. M., & Laran, J. (2016). The Effect of Stress on Consumer Saving and Spending. *Journal of Marketing Research*, *53*(5), 814–828. https://doi.org/10.1509/jmr.15.0319

Echterhoff, J., Liu, Y., Alessa, A., McAuley, J., & He, Z. (2024). *Cognitive Bias in Decision-Making with LLMs* (arXiv:2403.00811). arXiv. https://doi.org/10.48550/arXiv.2403.00811

Ethayarajh, K. (2019). *How Contextual are Contextualized Word Representations? Comparing the Geometry of BERT, ELMo, and GPT-2 Embeddings* (arXiv:1909.00512). arXiv. https://doi.org/10.48550/arXiv.1909.00512

Gadiraju, V., Kane, S., Dev, S., Taylor, A., Wang, D., Denton, E., & Brewer, R. (2023). "I wouldn't say offensive but…": Disability-Centered Perspectives on Large Language Models. *Proceedings of the 2023 ACM Conference on Fairness, Accountability, and Transparency*, 205–216. https://doi.org/10.1145/3593013.3593989

Gallagher, C. E., Watt, M. C., Weaver, A. D., & Murphy, K. A. (2017). "I fear, therefore, I shop!" exploring anxiety sensitivity in relation to compulsive buying. *Personality and Individual Differences*, *104*, 37–42.

Greshake, K., Abdelnabi, S., Mishra, S., Endres, C., Holz, T., & Fritz, M. (2023). Not What You've Signed Up For: Compromising Real-World LLM-Integrated Applications with Indirect Prompt Injection. *Proceedings of the 16th ACM Workshop on Artificial Intelligence and Security*, 79–90. CCS '23: ACM SIGSAC Conference on Computer and Communications Security. https://doi.org/10.1145/3605764.3623985

Hartley, C. A., & Phelps, E. A. (2012). Anxiety and decision-making. *Biological Psychiatry*, *72*(2), 113–118. https://doi.org/10.1016/j.biopsych.2011.12.027

He, J., Chen, S., Zhang, F., & Yang, Z. (2024). *From Words to Actions: Unveiling the Theoretical Underpinnings of LLM-Driven Autonomous Systems* (arXiv:2405.19883). arXiv. https://doi.org/10.48550/arXiv.2405.19883

Hill, D., Conner, M., Clancy, F., Moss, R., Wilding, S., Bristow, M., & O'Connor, D. B. (2022). Stress and eating behaviours in healthy adults: A systematic review and meta-analysis. *Health Psychology Review*, *16*(2), 280–304. https://doi.org/10.1080/17437199.2021.1923406

Jiang, L., Zhu, G., Sun, J., Cao, J., & Wu, J. (2025). Exploring the occupational biases and stereotypes of Chinese large language models. *Scientific Reports*, *15*(1), 18777. https://doi.org/10.1038/s41598-025-03893-w

Jin, H., Guo, J., Lin, Q., Wu, S., Hu, W., & Li, X. (2024). Comparative study of Claude 3.5-Sonnet and human physicians in generating discharge summaries for patients with renal insufficiency: Assessment of efficiency, accuracy, and quality. *Frontiers in Digital Health*, *6*, 1456911. https://doi.org/10.3389/fdgth.2024.1456911

Kamruzzaman, M., Shovon, M. M. I., & Kim, G. L. (2024). *Investigating Subtler Biases in LLMs: Ageism, Beauty, Institutional, and Nationality Bias in Generative Models* (arXiv:2309.08902). arXiv. https://doi.org/10.48550/arXiv.2309.08902

Kumar, N. (2025, June 19). Walmart Statistics 2025 – Numbers of Stores & Customers. *DemandSage*. https://www.demandsage.com/walmart-statistics/

Lazebnik, T., & Shami, L. (2025). *Investigating Tax Evasion Emergence Using Dual Large Language Model and Deep Reinforcement Learning Powered Agent-based Simulation* (arXiv:2501.18177). arXiv. https://doi.org/10.48550/arXiv.2501.18177

Lerner, J. S., Li, Y., Valdesolo, P., & Kassam, K. S. (2015). Emotion and Decision Making. *Annual Review of Psychology*, *66*(1), 799–823. https://doi.org/10.1146/annurev-psych-010213-115043

Lindström, A., Methnani, L., Krause, L., Ericson, P., de Rituerto de Troya, Í. M., Coelho Mollo, D., & Dobbe, R. (2025). Helpful, harmless, honest? Sociotechnical limits of AI alignment and safety through Reinforcement Learning from Human Feedback. *Ethics and Information Technology*, *27*(2), 28. https://doi.org/10.1007/s10676-025-09837-2

Liu, X., Yu, H., Zhang, H., Xu, Y., Lei, X., Lai, H., Gu, Y., Ding, H., Men, K., Yang, K., Zhang, S., Deng, X., Zeng, A., Du, Z., Zhang, C., Shen, S., Zhang, T., Su, Y., Sun, H., … Tang, J. (2023). *AgentBench: Evaluating LLMs as Agents* (arXiv:2308.03688). arXiv. https://doi.org/10.48550/arXiv.2308.03688

Loewenstein, G., Weber, E. U., Hsee, C. K., & Welch, N. (2001). Risk as feelings. *Psychological Bulletin*, *127*(2), 267–286. https://doi.org/10.1037/0033-2909.127.2.267

Macht, M. (2008). How emotions affect eating: A five-way model. *Appetite*, *50*(1), 1–11. https://doi.org/10.1016/j.appet.2007.07.002

Mikolov, T., Sutskever, I., Chen, K., Corrado, G. S., & Dean, J. (2013). Distributed representations of words and phrases and their compositionality. *Advances in Neural Information Processing Systems*, *26*. https://proceedings.neurips.cc/paper/2013/hash/9aa42b31882ec039965f3c4923ce901b-Abstract.html

Mitchell, M., & Krakauer, D. C. (2023). The debate over understanding in AI's large language models. *Proceedings of the National Academy of Sciences*, *120*(13), e2215907120. https://doi.org/10.1073/pnas.2215907120

Mittelstadt, B. (2019). Principles alone cannot guarantee ethical AI. *Nature Machine Intelligence*, *1*(11), 501–507. https://doi.org/10.1038/s42256-019-0114-4

Mozikov, M., Severin, N., Bodishtianu, V., Glushanina, M., Nasonov, I., Orekhov, D., Vladislav, P., Makovetskiy, I., Baklashkin, M., & Lavrentyev, V. (2024). EAI: Emotional decision-making of LLMs in strategic games and ethical dilemmas. *Advances in Neural Information Processing Systems*, *37*, 53969–54002.

Muthusamy, V., Rizk, Y., Kate, K., Venkateswaran, P., Isahagian, V., Gulati, A., & Dube, P. (2023). Towards large language model-based personal agents in the enterprise: Current trends and open problems. In H. Bouamor, J. Pino, & K. Bali (Eds.), *Findings of the Association for Computational Linguistics: EMNLP 2023* (pp. 6909–6921). Association for Computational Linguistics. https://doi.org/10.18653/v1/2023.findings-emnlp.461

Nadeem, M., Bethke, A., & Reddy, S. (2020). *StereoSet: Measuring stereotypical bias in pretrained language models* (arXiv:2004.09456). arXiv. https://doi.org/10.48550/arXiv.2004.09456

Nozza, D., Bianchi, F., Lauscher, A., & Hovy, D. (2022). Measuring harmful sentence completion in language models for LGBTQIA+ individuals. *Proceedings of the Second Workshop on Language Technology for Equality, Diversity and Inclusion*. https://doi.org/10.18653/v1/2022.ltedi-1.4

Olah, C., Cammarata, N., Schubert, L., Goh, G., Petrov, M., & Carter, S. (2020). Zoom in: An introduction to circuits. *Distill*, *5*(3), e00024-001.

OpenAI. (2025, August 7). *Introducing ChatGPT agent: Bridging research and action*. https://openai.com/index/introducing-chatgpt-agent/

Ouyang, L., Wu, J., Jiang, X., Almeida, D., Wainwright, C., Mishkin, P., Zhang, C., Agarwal, S., Slama, K., & Ray, A. (2022). Training language models to follow instructions with human feedback. *Advances in Neural Information Processing Systems*, *35*, 27730–27744.

OWASP. (2025). *OWASP Top 10 for LLM Apps & Gen AI Agentic Security Initiative*. OWASP. https://hal.science/hal-04985337

Park, J. S., O'Brien, J. C., Cai, C. J., Morris, M. R., Liang, P., & Bernstein, M. S. (2023). *Generative Agents: Interactive Simulacra of Human Behavior* (arXiv:2304.03442). arXiv. https://doi.org/10.48550/arXiv.2304.03442

Parrish, A., Chen, A., Nangia, N., Padmakumar, V., Phang, J., Thompson, J., Htut, P. M., & Bowman, S. R. (2022). *BBQ: A Hand-Built Bias Benchmark for Question Answering* (arXiv:2110.08193). arXiv. http://arxiv.org/abs/2110.08193

Perez, F., & Ribeiro, I. (2022). *Ignore Previous Prompt: Attack Techniques For Language Models* (arXiv:2211.09527). arXiv. https://doi.org/10.48550/arXiv.2211.09527

Pessoa, L. (2009). How do emotion and motivation direct executive control? *Trends in Cognitive Sciences*, *13*(4), 160–166.

Roer, G. E., Lien, L., Bolstad, I., Aaseth, J. O., & Abebe, D. S. (2023). The impact of PTSD on risk of cardiometabolic diseases: A national patient cohort study in Norway. *BMC Psychiatry*, *23*(1), 349. https://doi.org/10.1186/s12888-023-04866-x

Saxena, S. (2021). *Food Nutrition Dataset* [Dataset]. Kaggle. https://www.kaggle.com/datasets/shrutisaxena/food-nutrition-dataset

Schick, T., Dwivedi-Yu, J., Dessì, R., Raileanu, R., Lomeli, M., Hambro, E., Zettlemoyer, L., Cancedda, N., & Scialom, T. (2023). Toolformer: Language models can teach themselves to use tools. *Advances in Neural Information Processing Systems*, *36*, 68539–68551.

Schwabe, L., Tegenthoff, M., Höffken, O., & Wolf, O. T. (2010). Concurrent glucocorticoid and noradrenergic activity shifts instrumental behavior from goal-directed to habitual control. *Journal of Neuroscience*, *30*(24), 8190–8196.

Schwabe, L., & Wolf, O. T. (2009). Stress prompts habit behavior in humans. *Journal of Neuroscience*, *29*(22), 7191–7198.

Sclar, M., Choi, Y., Tsvetkov, Y., & Suhr, A. (2024). *Quantifying Language Models' Sensitivity to Spurious Features in Prompt Design or: How I learned to start worrying about prompt formatting* (arXiv:2310.11324). arXiv. https://doi.org/10.48550/arXiv.2310.11324

Sorin, V., Brin, D., Barash, Y., Konen, E., Charney, A., Nadkarni, G., & Klang, E. (2024). Large Language Models and Empathy: Systematic Review. *Journal of Medical Internet Research*, *26*, e52597. https://doi.org/10.2196/52597

Spielberger, C. D. (1983). *State-trait anxiety inventory for adults*. https://psycnet.apa.org/doiLanding?doi=10.1037%2Ft06496-000

Stefanovics, E. A., Potenza, M. N., & Pietrzak, R. H. (2020). PTSD and obesity in US military veterans: Prevalence, health burden, and suicidality. *Psychiatry Research*, *291*, 113242.

Tamkin, A., Askell, A., Lovitt, L., Durmus, E., Joseph, N., Kravec, S., Nguyen, K., Kaplan, J., & Ganguli, D. (2023). *Evaluating and Mitigating Discrimination in Language Model Decisions* (arXiv:2312.03689). arXiv. http://arxiv.org/abs/2312.03689

Tomiyama, A. J. (2019). Stress and Obesity. *Annual Review of Psychology*, *70*(1), 703–718. https://doi.org/10.1146/annurev-psych-010418-102936

Torres, S. J., & Nowson, C. A. (2007). Relationship between stress, eating behavior, and obesity. *Nutrition (Burbank, Los Angeles County, Calif.)*, *23*(11–12), 887–894. https://doi.org/10.1016/j.nut.2007.08.008

Turk, V. (2025). Who bought this smoked salmon? How 'AI agents' will change the internet (and shopping lists). *The Guardian*. https://www.theguardian.com/technology/2025/mar/09/who-bought-this-smoked-salmon-how-ai-agents-will-change-the-internet-and-shopping-lists

UK Department of Health. (2011). *Nutrient Profiling Technical Guidance*. UK Department of Health. https://www.gov.uk/government/publications/the-nutrient-profiling-model

van der Bend, D. L. M., van Eijsden, M., van Roost, M. H. I., de Graaf, K., & Roodenburg, A. J. C. (2022). The Nutri-Score algorithm: Evaluation of its validation process. *Frontiers in Nutrition*, *9*. https://doi.org/10.3389/fnut.2022.974003

Venkit, P., Gautam, S., Panchanadikar, R., Huang, T.-H. "Kenneth," & Wilson, S. (2023). *Nationality Bias in Text Generation* (arXiv:2302.02463). arXiv. https://doi.org/10.48550/arXiv.2302.02463

Verma, M., Bhambri, S., & Kambhampati, S. (2024). *On the Brittle Foundations of ReAct Prompting for Agentic Large Language Models* (arXiv:2405.13966). arXiv. https://doi.org/10.48550/arXiv.2405.13966

Wang, Q., Wang, Z., Su, Y., Tong, H., & Song, Y. (2024). *Rethinking the Bounds of LLM Reasoning: Are Multi-Agent Discussions the Key?* (arXiv:2402.18272). arXiv. https://doi.org/10.48550/arXiv.2402.18272

Wu, X., Wang, H., Yan, Z., Tang, X., Xu, P., Siok, W.-T., Li, P., Gao, J.-H., Lyu, B., & Qin, L. (2025). *AI shares emotion with humans across languages and cultures* (arXiv:2506.13978). arXiv. https://doi.org/10.48550/arXiv.2506.13978

Zao-Sanders, M. (2025). How People Are Really Using Gen AI in 2025. *Harvard Business Review*. https://hbr.org/2025/04/how-people-are-really-using-gen-ai-in-2025

Zhang, C., Zhang, C., Zheng, S., Qiao, Y., Li, C., Zhang, M., Dam, S. K., Thwal, C. M., Tun, Y. L., Huy, L. L., kim, D., Bae, S.-H., Lee, L.-H., Yang, Y., Shen, H. T., Kweon, I. S., & Hong, C. S. (2023). *A Complete Survey on Generative AI (AIGC): Is ChatGPT from GPT-4 to GPT-5 All You Need?* (arXiv:2303.11717). arXiv. https://doi.org/10.48550/arXiv.2303.11717

Zhou, L., Schellaert, W., Martínez-Plumed, F., Moros-Daval, Y., Ferri, C., & Hernández-Orallo, J. (2024). Larger and more instructable language models become less reliable. *Nature*, *634*(8032), 61–68. https://doi.org/10.1038/s41586-024-07930-y

Zhou, S., Xu, F. F., Zhu, H., Zhou, X., Lo, R., Sridhar, A., Cheng, X., Ou, T., Bisk, Y., Fried, D., Alon, U., & Neubig, G. (2024). *WebArena: A Realistic Web Environment for Building Autonomous Agents* (arXiv:2307.13854). arXiv. https://doi.org/10.48550/arXiv.2307.13854

# Inducing State Anxiety in LLM Agents Reproduces Human-Like Biases in Consumer Decision-Making

*Supplementary Information*

**Supplemental Table 1.** Within-condition changes in Basket Health Scores (BHS) across all experimental conditions. *(pages 2-6)*

**Supplemental Table 2.** Budget-level effects of traumatic prompts on Basket Health Scores (BHS). *(page 7)*

**Supplemental Table 3.** Model-level effects of traumatic prompts on Basket Health Scores (BHS). *(page 8)*

**Supplemental Table 4.** Model- and budget-level effects of traumatic prompts on Basket Health Scores (BHS). *(pages 9-10)*

**Supplemental Table 1. Within-condition changes in Basket Health Scores (BHS) across all experimental conditions.** Each row reports descriptive and inferential statistics for a single condition defined by the LLM model (ChatGPT-5, Gemini 2.5, Claude 3.5-Sonnet), budget ($27, $54, $108), and traumatic prompt (Accident, Ambush, Disaster, Interpersonal Violence, Military). Reported values include mean and SD of the change in BHS (Δ = post – pre), effect sizes (Cohen's *d*), raw p-values, and FDR-adjusted p-values. All tests were conducted at the run level (n = 50 per condition).

**Within-condition changes in BHS across all experimental conditions**

**45 Unique Conditions (3 LLMs * 3 Budgets * 5 Traumatic Narratives)**

| LLM | Budget | Prompt | n | Mean Δ | SD Δ | Cohen's *d* | p-value | p-value (FDR) |
|---|---|---|---|---|---|---|---|---|
| ChatGPT 5 | 27 | accident | 50 | −0.117 | 0.084 | −1.401 | 0.0000 | 0.0000 |
| ChatGPT 5 | 27 | ambush | 50 | −0.111 | 0.072 | −1.552 | 0.0000 | 0.0000 |
| ChatGPT 5 | 27 | disaster | 50 | −0.061 | 0.045 | −1.360 | 0.0000 | 0.0000 |
| ChatGPT 5 | 27 | interpersonal | 50 | −0.092 | 0.079 | −1.170 | 0.0000 | 0.0000 |
| ChatGPT 5 | 27 | military | 50 | −0.092 | 0.061 | −1.510 | 0.0000 | 0.0000 |
| ChatGPT 5 | 54 | accident | 50 | −0.149 | 0.077 | −1.945 | 0.0000 | 0.0000 |
| ChatGPT 5 | 54 | ambush | 50 | −0.123 | 0.066 | −1.864 | 0.0000 | 0.0000 |
| ChatGPT 5 | 54 | disaster | 50 | −0.097 | 0.046 | −2.087 | 0.0000 | 0.0000 |
| ChatGPT 5 | 54 | interpersonal | 50 | −0.057 | 0.089 | −0.641 | 0.0000 | 0.0000 |
| ChatGPT 5 | 54 | military | 50 | −0.082 | 0.065 | −1.260 | 0.0000 | 0.0000 |
| ChatGPT 5 | 108 | accident | 50 | −0.114 | 0.052 | −2.212 | 0.0000 | 0.0000 |
| ChatGPT 5 | 108 | ambush | 50 | −0.118 | 0.069 | −1.704 | 0.0000 | 0.0000 |
| ChatGPT 5 | 108 | disaster | 50 | −0.090 | 0.064 | −1.406 | 0.0000 | 0.0000 |
| ChatGPT 5 | 108 | interpersonal | 50 | −0.071 | 0.103 | −0.687 | 0.0000 | 0.0000 |
| ChatGPT 5 | 108 | military | 50 | −0.103 | 0.050 | −2.071 | 0.0000 | 0.0000 |
| Claude 3.5 Sonnet | 27 | accident | 50 | −0.136 | 0.068 | −2.010 | 0.0000 | 0.0000 |
| Claude 3.5 Sonnet | 27 | ambush | 50 | −0.149 | 0.027 | −5.583 | 0.0000 | 0.0000 |

| Model | | | | | | | | |
|---|---|---|---|---|---|---|---|---|
| Claude 3.5 Sonnet | 27 | disaster | 50 | −0.124 | 0.024 | −5.067 | 0.0000 | 0.0000 |
| Claude 3.5 Sonnet | 27 | interpersonal | 50 | −0.070 | 0.046 | −1.523 | 0.0000 | 0.0000 |
| Claude 3.5 Sonnet | 27 | military | 50 | −0.125 | 0.040 | −3.143 | 0.0000 | 0.0000 |
| Claude 3.5 Sonnet | 54 | accident | 50 | −0.115 | 0.028 | −4.180 | 0.0000 | 0.0000 |
| Claude 3.5 Sonnet | 54 | ambush | 50 | −0.113 | 0.065 | −1.727 | 0.0000 | 0.0000 |
| Claude 3.5 Sonnet | 54 | disaster | 50 | −0.077 | 0.039 | −1.952 | 0.0000 | 0.0000 |
| Claude 3.5 Sonnet | 54 | interpersonal | 50 | −0.104 | 0.069 | −1.505 | 0.0000 | 0.0000 |
| Claude 3.5 Sonnet | 54 | military | 50 | −0.087 | 0.033 | −2.676 | 0.0000 | 0.0000 |
| Claude 3.5 Sonnet | 108 | accident | 50 | −0.128 | 0.037 | −3.487 | 0.0000 | 0.0000 |
| Claude 3.5 Sonnet | 108 | ambush | 50 | −0.143 | 0.044 | −3.268 | 0.0000 | 0.0000 |
| Claude 3.5 Sonnet | 108 | disaster | 50 | −0.050 | 0.026 | −1.957 | 0.0000 | 0.0000 |
| Claude 3.5 Sonnet | 108 | interpersonal | 50 | −0.090 | 0.064 | −1.405 | 0.0000 | 0.0000 |
| Claude 3.5 Sonnet | 108 | military | 50 | −0.115 | 0.049 | −2.367 | 0.0000 | 0.0000 |
| Gemini 2.5 | 27 | accident | 50 | −0.131 | 0.072 | −1.831 | 0.0000 | 0.0000 |
| Gemini 2.5 | 27 | ambush | 50 | −0.149 | 0.053 | −2.796 | 0.0000 | 0.0000 |
| Gemini 2.5 | 27 | disaster | 50 | −0.088 | 0.046 | −1.913 | 0.0000 | 0.0000 |
| Gemini 2.5 | 27 | interpersonal | 50 | −0.111 | 0.061 | −1.816 | 0.0000 | 0.0000 |
| Gemini 2.5 | 27 | military | 50 | −0.111 | 0.064 | −1.743 | 0.0000 | 0.0000 |
| Gemini 2.5 | 54 | accident | 50 | −0.122 | 0.072 | −1.698 | 0.0000 | 0.0000 |
| Gemini 2.5 | 54 | ambush | 50 | −0.114 | 0.075 | −1.528 | 0.0000 | 0.0000 |

| Gemini 2.5 | 54  | disaster      | 50 | −0.093 | 0.077 | −1.204 | 0.0000 | 0.0000 |
| Gemini 2.5 | 54  | interpersonal | 50 | −0.105 | 0.075 | −1.396 | 0.0000 | 0.0000 |
| Gemini 2.5 | 54  | military      | 50 | −0.122 | 0.047 | −2.616 | 0.0000 | 0.0000 |
| Gemini 2.5 | 108 | accident      | 50 | −0.113 | 0.091 | −1.244 | 0.0000 | 0.0000 |
| Gemini 2.5 | 108 | ambush        | 50 | −0.113 | 0.054 | −2.105 | 0.0000 | 0.0000 |
| Gemini 2.5 | 108 | disaster      | 50 | −0.129 | 0.066 | −1.957 | 0.0000 | 0.0000 |
| Gemini 2.5 | 108 | interpersonal | 50 | −0.031 | 0.050 | −0.614 | 0.0000 | 0.0000 |
| Gemini 2.5 | 108 | military      | 50 | −0.099 | 0.065 | −1.534 | 0.0000 | 0.0000 |

**Supplemental Table 2. Budget-level effects of traumatic prompts on Basket Health Scores (BHS).** Each row reports descriptive and inferential statistics for pooled conditions defined by budget constraints across all LLM models and all anxiety-inducing traumatic prompts. Reported values include mean and SD of the change in BHS (Δ = post – pre), effect sizes (Cohen's *d*), raw p-values, and FDR-adjusted p-values.

**Stratified Analysis by Budget Condition**

**Pooled across all anxiety-inducing prompts and LLMs**

| Budget | n | Mean Δ | SD Δ | Cohen's *d* | 95% CI Lower | 95% CI Upper | p-value | p-value (FDR) |
|---|---|---|---|---|---|---|---|---|
| **Low** | 750 | −0.111 | 0.063 | −1.754 | −0.116 | −0.107 | 0.0000 | 0.0000 |
| **Medium** | 750 | −0.104 | 0.067 | −1.550 | −0.109 | −0.099 | 0.0000 | 0.0000 |
| **High** | 750 | −0.100 | 0.068 | −1.481 | −0.105 | −0.096 | 0.0000 | 0.0000 |

**Supplemental Table 3. Model-level effects of traumatic prompts on Basket Health Scores (BHS).** Each row reports descriptive and inferential statistics for pooled conditions defined by LLM models across all budget constraints and all anxiety-inducing traumatic prompts. Reported values include mean and SD of the change in BHS (Δ = post – pre), effect sizes (Cohen's *d*), raw p-values, and FDR-adjusted p-values.

**Stratified Analysis by LLM Model**

**Pooled across all anxiety-inducing prompts and budgets**

| LLM | n | Mean Δ | SD Δ | Cohen's *d* | 95% CI Lower | 95% CI Upper | p-value | p-value (FDR) |
|---|---|---|---|---|---|---|---|---|
| **ChatGPT 5** | 750 | −0.098 | 0.073 | −1.344 | −0.104 | −0.093 | 0.0000 | 0.0000 |
| **Claude 3.5 Sonnet** | 750 | −0.108 | 0.054 | −2.022 | −0.112 | −0.105 | 0.0000 | 0.0000 |
| **Gemini 2.5** | 750 | −0.109 | 0.070 | −1.557 | −0.114 | −0.104 | 0.0000 | 0.0000 |

**Supplemental Table 4. Model- and budget-level effects of traumatic prompts on Basket Health Scores (BHS).** Each row reports descriptive and inferential statistics for pooled conditions defined by LLM model (ChatGPT-5, Gemini 2.5, Claude 3.5-Sonnet) or budget constraint ($27, $54, $108), across all five anxiety-inducing traumatic prompts. Reported values include mean and SD of the change in BHS (Δ = post – pre), effect sizes (Cohen's *d*), raw p-values, and FDR-adjusted p-values.

**LLM * Budget Interaction Analysis**

**Pooled across all traumatic narratives**

| LLM | Budget | n | Mean Δ | SD Δ | p-value | Cohen's *d* | p-value (FDR) |
|---|---|---|---|---|---|---|---|
| ChatGPT 5 | 27 | 250 | −0.095 | 0.071 | 0.0000 | −1.324 | 0.0000 |
| ChatGPT 5 | 54 | 250 | −0.102 | 0.076 | 0.0000 | −1.328 | 0.0000 |
| ChatGPT 5 | 108 | 250 | −0.099 | 0.072 | 0.0000 | −1.382 | 0.0000 |
| Claude 3.5 Sonnet | 27 | 250 | −0.121 | 0.051 | 0.0000 | −2.362 | 0.0000 |
| Claude 3.5 Sonnet | 54 | 250 | −0.099 | 0.052 | 0.0000 | −1.921 | 0.0000 |
| Claude 3.5 Sonnet | 108 | 250 | −0.105 | 0.056 | 0.0000 | −1.886 | 0.0000 |

| | | | | | | |
|---|---|---|---|---|---|---|
| **Gemini 2.5** | 27  | 250 | −0.118 | 0.063 | 0.0000 | −1.879 | 0.0000 |
| **Gemini 2.5** | 54  | 250 | −0.111 | 0.070 | 0.0000 | −1.581 | 0.0000 |
| **Gemini 2.5** | 108 | 250 | −0.097 | 0.075 | 0.0000 | −1.302 | 0.0000 |